\documentstyle[prl,aps,preprint,tighten,floats,aps]{revtex}

\def\be{\begin{equation}}
\def\ee{\end{equation}}
\def\bear{\begin{eqnarray}}
\def\eear{\end{eqnarray}}
\def\beqn{\begin{eqnarray}}
\def\eeqn{\end{eqnarray}}

\def\beq{\begin{equation} }
\def\eeq{\end{equation} }

\def\ben{\begin{eqnarray} }
\def\een{\end{eqnarray} }

\def\mod#1{{\rm (mod~2)} }

\def\Tr{{\rm Tr}}

\def\Tr{{\rm Tr}\,}

\begin{document}
\draft
\preprint{\vbox{\baselineskip=12pt
\rightline{UPR-0810-T}
\vskip0.2truecm
\rightline{hep-ph/9808321}}}
\title{Units and Numerical Values of the  Effective Couplings in
Perturbative Heterotic String Vacua} 
\author{ Mirjam Cveti\v c${}^*$${}^{\dagger}$, Lisa Everett${}^{\dagger}$,
and Jing Wang${}^{\dagger}$}
\address{${}^{\dagger}$Department of Physics and Astronomy \\ 
University of Pennsylvania, Philadelphia PA 19104-6396, USA \\
${}^*$Institute for Theoretical Physics\\
University of California, Santa Barbara, CA 93106, USA}
\maketitle
\begin{abstract}
We determine the units and numerical values for a class of 
couplings in the  effective theory  of perturbative heterotic string
vacua, with
the emphasis  on the correct translation between the
 canonical
 gauge coupling $g$    and Planck scale $M_{Planck}\sim 1.2 \times 10^{19}$ GeV 
 as  used in the effective theory description and 
 the string coupling
 $g_{string}$    and string tension $\alpha'$ as used in the $S$-matrix 
 amplitude calculation. 
In particular, we  determine the effective couplings in the superpotential
 and  revisit the Fayet-Iliopoulos (FI) term  in a class of models with an 
 anomalous $U(1)$.  We  derive the  values
  of the effective  Yukawa couplings   (at the third and fourth
 order)  after the
restabilization of vacuum  along a particular $F$- and $D$-flat
direction and show that they are {\it comparable} in magnitude. 
The  result corrects  results quoted in the literature,  and may have
implications for  the string derived phenomenology, e.g., that  of
fermion textures.
 \end{abstract}
\newpage
\noindent{\it Introduction.}
For a large class of perturbative string vacua the 
string calculations, exact at  at the tree level of string perturbation,  
can
be done to determine couplings in the effective Lagrangian.  
In particular, the techniques to determine the superpotential terms   have been
developed. A class of 
 quasi-realistic perturbative heterotic string models, based on free fermionic
 construction,  constitute a set of  such models. 
 However, these models  in general  also contain 
anomalous $U(1)$ \cite{CHL,FNY1,AF1,NAHE,AFT,genanoms} with   the nonzero
Fayet-Iliopoulos (FI) $D$- term generated at the one-genus
order  \cite{DSW,ADS,DKI,AS}.  
In order to maintain the $D$- and $F$- flatness, certain scalar fields 
acquire vacuum expectation values (VEV's) 
leading to a shifted (``restabilized")
supersymmetric string vacuum.  (For a recent approach to
systematically analyze such flat directions see \cite{cceel2,cceel3}.)

In general, the vacuum shift reduces the rank of the gauge group, and
generates effective mass terms and effective trilinear couplings from 
higher-order terms in the superpotential after replacing the fields in the
flat direction by their VEV's. (For a recent
 calculation of the mass spectrum and effective couplings of a prototype
free fermionic string model see \cite{cceelw}, and for the study of the 
effects of the decoupling of massive states  see \cite{ceew}.)  
The magnitudes of the effective couplings generated in this way depend
on the values of the VEV's of the fields in the flat direction, which are
set by the FI term~\footnote{In particular, while nonzero VEV's  of the
fields that restabilize the vacuum may depend on free
parameters (moduli) $(\psi_i)$, their maximum values are  determined
by the magnitude  of the FI-term. One could view moduli parameters as lying on
a ``hyper-sphere'' with radius-square  proportional to  the magnitude of
FI-term.}. 

The effective trilinear superpotential couplings  of the surviving massless
spectrum along
a particular flat direction may have important implications for the
string derived phenomenology of the model and thus the precise determination
of such effective  couplings in the model has consequences for the low energy
physics of the model. (See ~\cite{cceelw} for recent discussions.)
Therefore, it is important that the numerical determination
of  the couplings in the effective theory is correct.  In this note we
provide a dictionary that properly translates between the  $S$-matrix
calculations in  the perturbative heterotic string theory (and units used
there) with the corresponding  terms  and units used  in the effective
theory. 

We focus on the proper determination of numerical values and units 
of the coefficients of the third and higher-order terms of the
superpotential and of the FI term, which are both calculable in perturbative
heterotic
string theory~ \footnote{However,
nonrenormalizable terms competitive in strength are also present in the
original superpotential, as well as generated in a number of other ways,
such as via the decoupling of heavy states\cite{ceew}, a nonminimal K\"{a}hler
potential, and the corrections to the K\"{a}hler potential due to the large
VEV's.}. We find that it is important to make the proper
identification of the gauge coupling in the effective theory which is
consistent with the canonical definition in particle physics as well as
the proper restoration of the $\alpha'$ (string tension) unit in the 
computation of physical quantities, since  these issues are a source of 
confusion in the literature.  
We illustrate the results with a prototype string
model (Model 5 of \cite{CHL}), for which the flat directions and
effective couplings have been analyzed \cite{cceel2,cceelw}, as well as
 some of the models of the NAHE type \cite{NAHE}. 

\noindent{\it String  Amplitude Calculations.}
In string theory, the  effective Lagrangian is determined via string  
$S$-matrix amplitude calculations.  For each physical state $\Phi_i$ in 
the theory
there is a string vertex operator $V_{\Phi_i}$, whose structure is fully 
determined from the world-sheet constraints of the string theory. Conformal
invariance provides the rules for calculating such string
amplitudes~\cite{SF,GSW}.  In particular, the string amplitude
  $A_n^{(k)}$ for a  ``scattering'' of
   $n$-states $\{ \Phi_1,\cdots , \Phi_n\}$ at genus-$k$
  of the theory can be schematically written as:
  \begin{equation}
  A_n^{(k)}= g_{string}^{n-2+2k} \langle V_{\phi_1}\cdots V_{\phi_n} \rangle \,
  \label{amplitude}
  \end{equation}
  in which $g_{string}$ is the string coupling, and the  amplitude $ 
\langle \cdots \rangle$  involves non-trivial world-sheet integrals $1/{2\pi\alpha'}\int d^2 z$ 
and the integral over the moduli of the Riemann world-sheet surfaces.   
In most calculations the string tension $\alpha '$ is set to  a
  constant (e.g., 2 or 1/2), but  can be restored  by properly identifying the
mass  dimension of the  term in the effective theory,  probed by the
string amplitude (\ref{amplitude}).
  
  The calculations of the three-graviton vertex  amplitude and the  three-point
   amplitude of the  non-Abelian gauge fields determines $M_{Planck}$ and the
   gauge coupling $g$ in the effective theory in terms of  $g_{string}$  and 
   $\alpha'$. (For ten-dimensional  heterotic string theory  we refer the 
reader to e.g., \cite{GHMR},
and for four-dimensional string theory see e.g., \cite{PG,GM,LT}.)
The result is the following:
\begin{equation}
g= \sqrt{2} g_{string}, \ \ \ 
M_{Planck}^2= \frac{16\pi}{g_{string}^2\alpha'}.
\label{gM}
\end{equation}
  The discrepancy of $\sqrt 2$ in the definition of the gauge coupling in
the  effective theory, often a source of confusion in the literature,  arises
due to the conventional definition of the gauge
coupling in the effective theory, which follows the convention that ${\rm
Tr} T^aT^b= 1/2\delta_{ab}, \
  \delta_{ab}$ for the fields in the fundamental representation of the
$SU(N)$ and $SO(N)$, respectively.  On the other hand, in string theory the vertex operators
for the gauge bosons are determined using the convention   that the
lattice length of the  field in the fundamental representation of $SO(N)$
is 2. (For further discussion of this issue see \cite{kap}.) 

\noindent{\it Effective Superpotential.} 
The terms in the effective superpotential are
can be easily determined from the  string amplitude calculation, by 
studying the following tree-level amplitude 
(at genus-zero)~\cite{cvetic,worldsheet,nano}:
 \begin{eqnarray}
A_n^{(0)}&=& g_{string}^{n-2}<V_{F_1}V_{F_2} V_{B_3} \cdots V_{B_n}>\nonumber\\
&=& g_{string}^{n-2} \left({1\over{2\pi}}\right)^{n-3}\int d^2z_1\cdots d^2z_{n-3} 
(V_{-\frac{1}{2}})_1 (V_{-\frac{1}{2}})_2
(V_{-1})_3 (V_{0})_4\cdots (V_{0})_n,
\label{superpot}
\end{eqnarray}
where the vertices $V_{F,~ B}$ refer to the  vertex operators for the 
fermionic- and bosonic- components of the chiral superfields,
respectively, and the subscripts $-{1\over 2}, \ {-1},  \ 0$ to the
corresponding ``conformal ghost pictures''.  The convention $\alpha' ={1\over 2}$
was taken in the second line.  The $S$-matrix amplitude  directly probes
the  corresponding terms in the superpotential. i.e. there are {\it no}
contributions  to the amplitude from the massless exchanges~\cite{cvetic}. (In
addition, the non-renormalization theorem \cite{DSWW} ensures that this
tree-level result is true to all orders in the string (genus) perturbative
expansion.)~\footnote{On the other hand the calculation of the  K\"ahler potential
is more involved; there are higher genus corrections, and
the contributions  of massless field exchanges have to be subtracted
in the corresponding string amplitudes. Note also that
after  vacuum restabilization (see later) higher order terms in the K\"ahler
potential may introduce significant corrections  to the  prefactor in the kinetic
energy of the surviving  massless fields and thus  they may change  quantitatively
the physical value of the corresponding effective Yukawa couplings. The discussion
of these effects is beyond the scope of the paper.}
 The amplitude (\ref{superpot} ) thus  directly determines 
the  $n$-th order coupling in the superpotential:
\begin{equation}
{\cal A}_n= g_{string}^{n-2} \left (\frac{{\sqrt{2
\alpha'}}}{2 \pi}\right)^{n-3} C_{n-3} I_{n-3}.
\label{coeff}
\end{equation}
(For simplicity of the notation, we suppress the superscript $0$ for
$A_{n}^{(0)}$.) Here, $C_{n-3}$ is the coefficient of
${\cal O} (1)$ which includes the
target space gauge group Clebsch-Gordan coefficients and the renormalization
factors of vetex operators (see e.g., \cite{LT}). The worldsheet integral
$I_{n-3}$ is of the form   
\begin{equation}
I_{n-3}=\int \Pi_{i=1}^{n-3} d^2z_i f(z_i,{\bar z}_i),
\end{equation}
where the function $f$ depends on the world-sheet
coordinates $z_i, \ {\bar z}_i$, as determined by the operator product
expansion of the operators that specify the string  vertices of the 
the physical states. In particular, for the free fermionic models
 and  $n=4$
the  typical worldsheet integral  $I_1$ is of the form (see e.g., \cite{nano}):
\begin{equation}
I_1=\int d^2 z  |z|^{N_1}|1-z|^{N_2}(1+ |z|+|z-1|)^{N_3}.
\label{fourth}
\end{equation}
For
free fermionic constructions (based on $Z_2\times Z_2$)  the allowed powers
$N_{1,2,3}$   in (\ref{fourth})  are  fractions  which are integer multiples of
$1/4$ and    their  values are constrained, yielding finite values for 
 $I_1$.  (The third factor in (\ref{fourth}) 
arises due to the contribution of
Ising fields with the exponent $N_3=\textstyle {1\over 2}$.)
 A typical example corresponds to the case $N_1=- 1$, $N_2=-{7\over 4}$,
 $N_3={1\over 2}$, thus yielding for $I_1$ in (\ref{fourth}) $I_1\sim  70$, which is
 close to saturating the lower bound $I_{min}\sim 60 $ corresponding to $N_1=-{5\over 4}$, $N_2=-{5\over 4}$,
 $N_3=0$. Analogously,  for $n=5$, a typical value for  $I_2$ is $\sim 400$. (See, e.g.,
\cite{nano,cceel1}).

Using the relationship between $\alpha'$, $g_{string}$, and $M_{Planck}$,
(\ref{coeff}) can be rewritten as follows:
\begin{equation}
\label{coeff1}
{\cal A}_n=  \frac{g}{\sqrt{2}} \left( \sqrt{\frac{8}{\pi}}
\right)^{n-3}\frac{C_{n-3}I_{n-3}}{M_{Planck}^{n-3}}.
\end{equation} 
The value of the amplitude ${\cal A}_3$ is given by ${\cal A}_3=C_0 
g_{string}$, in which  $C_0$ takes the values $\{\sqrt{2}, 1,
1/\sqrt{2}\}$~\cite{nano}.  The usual case is that $C_0=\sqrt{2}$, which
corresponds to the situation in which the string
vertex operators do not involve additional (real) world-sheet fermion
fields.  Thus, in this case
\begin{equation}
{\cal A}_3=\sqrt{2}g_{string}=g.
\end{equation} 

The value of the amplitude ${\cal A}_4$ is obtained from (\ref{coeff1}): 
\begin{equation}
{\cal A}_4=g\frac{ 2 C_1I_1}{\sqrt{\pi}M_{Planck}},
\end{equation}
with the explicit value depending on the value of $I_1$ (in  (\ref{fourth})).

\noindent{\it Fayet-Iliopoulos $D$- term.}
The presence of the anomalous $U(1)_A$ leads to the generation of a nonzero
Fayet-Iliopoulos term $\xi$ in the corresponding $D$- term\footnote{Our convention
for defining $D_{A}$ is that the corresponding $D$- term in the Lagrangian is 
$\frac{1}{2k_A}g^2D_{A}^2$.}
\begin{equation}
D_A=D_A^{(0)}+\xi
\label{xidef}
\end{equation}
at genus-one in string theory.
The FI term is obtained in \cite{ADS} by the 
genus-one calculation of the mass terms ($n=2$, $k=1$ in
(\ref{amplitude})), and is obtained in \cite{AS}
by the calculation of a dilaton tadpole (which is a genus-two
effect). The FI term can also be determined just based on anomaly cancellation
arguments at the effective theory level \cite{DSW}. In dimensionless
units (setting $\alpha'=2$), the result is~\cite{ADS}:
\begin{equation}
\xi=\frac{\Tr Q_A}{192\pi^2}.
\label{xidiml}
\end{equation}
After properly restoring the mass units, the FI term is 
\begin{equation}
\xi =\frac{\Tr Q_A}{192\pi^2}\frac{2}{\alpha'}
= \frac{g^2 \Tr Q_A}{192\pi^2}\frac{M^2_{Planck}}{16\pi}.
\label{fid}
\end{equation}
After vacuum restablization, i.e., ensuring $D_{A}=0$, certain fields 
acquire VEV's. For notational simplicity, we write 
$D_{A}^{(0)}=Q_{A}^{(0)}|\langle \phi \rangle |^{2}=-\xi$. Here, $\langle 
\phi \rangle$ denotes  a typical VEV for the fields involved in the flat 
direction, and $Q_{A}^{(0)}$ denotes an effective anomalous $U(1)$ 
charge. (See \cite{cceel2} for a more detailed discussion.) 

\noindent{\it Examples and Summary.}
To illustrate the results stated above, we turn to the calculation of the
effective trilinear couplings in a prototype string model (Model 5 in
Ref.~\cite{CHL}, which is also
addressed in \cite{cceelw}).  
In this model, with the normalization of the $U(1)$ charges as listed in
\cite{CHL,cceel2}, ${\rm Tr} Q_A=-1536$, and thus the FI term is
determined from (\ref{fid})
to be $\xi=-0.0103~M_{Planck}^{2}$. (The FI term depends on the value of $g_{string}$ at
the string (unification) scale $M_{String} \sim 5 \times 10^{17}\,GeV$; 
to be consistent with the analysis in \cite{cceelw}, we choose 
$g_{string}\sim 0.56$, such that $g\sim 0.80$.)  

In general, the VEV's depend on the ratio $\sqrt{\xi/Q_A^{(0)}}$, in which
$Q_A^{(0)}$ is as previously defined. 
The VEV's of the fields of the most general flat direction of the
subset considered have been presented in \cite{cceelw}, with the result 
that the typical VEV of a given field $\phi$ in the flat direction has 
$Q_{A}^{(0)}=64$, and hence $\langle \phi
\rangle=\sqrt{-\xi}/8=0.0127~M_{Planck}$.   

The trilinear couplings of the original superpotential generically have the 
coupling strengths $\sqrt{2} g_{string}=g\sim 0.80$.  Therefore, the 
ratio of the effective trilinear couplings generated from the 
fourth-order terms in the
superpotential to the original trilinear terms is (taking $C_1=1$, 
and $I_1 \sim 70$):
\begin{equation}
\label{ratio}
\frac{W_4}{W_3}=\left( \frac{ 2g I_1\langle \phi\rangle}{\sqrt{\pi}M_{Planck}} 
\right) \frac{1}{g}=1.003,
\end{equation}
 such that in this model, the effective trilinear couplings generated from
{\it fourth-order terms  are  competitive
with the original  trilinear terms}, contrary  to the expectations based 
on naive estimates.  Inspection of  the effective trilinear terms  arising  
at  the fifth order (as well as at higher orders)  shows that they are 
indeed   suppressed relative to the terms in $W_3$; for example, 
with $C_2=1$ and $I_2\sim 400$, one obtains $ {W_5}/{W_3}\sim 0.1$. 

Another example is  within the NAHE models~\cite{NAHE}  along  
particular flat directions.  The value $ {\rm Tr} Q_A/Q_A^{(0)} = - 12$ is 
typical for a set of flat directions. (See,   e.g.,  \cite{cceel3} for 
a discussion of this class of flat directions in these models.)  Thus, in 
this case a typical VEV is 
$\langle\phi\rangle= \sqrt{-\xi/Q_A^{(0)}}\sim 0.0090\, M_{Planck}$, such that the 
effective trilinear couplings generated from the fourth-order terms are 
again comparable to the elementary trilinear term; for example,  with 
$C_0=\sqrt{2}$,
$C_1=1$, and $I_1\sim 70$, one obtains $W_4/W_3\sim 0.71$. 

\acknowledgments
This work was supported in part by U.S. Department of Energy Grant No. 
DOE-EY-76-02-3071 and NSF  grant PHY94-07194 (MC).  We would like to thank
 K. Dienes and  V. Kaplunovsky for useful correspondence, and G. 
Cleaver,  J. R. Espinosa,  and especially P. Langacker for  many discussions.
\newpage

\def\B#1#2#3{\/ {\bf B#1} (19#2) #3}
\def\NPB#1#2#3{{\it Nucl.\ Phys.}\/ {\bf B#1} (19#2) #3}
\def\PLB#1#2#3{{\it Phys.\ Lett.}\/ {\bf B#1} (19#2) #3}
\def\PRD#1#2#3{{\it Phys.\ Rev.}\/ {\bf D#1} (19#2) #3}
\def\PRL#1#2#3{{\it Phys.\ Rev.\ Lett.}\/ {\bf #1} (19#2) #3}
\def\PRT#1#2#3{{\it Phys.\ Rep.}\/ {\bf#1} (19#2) #3}
\def\MODA#1#2#3{{\it Mod.\ Phys.\ Lett.}\/ {\bf A#1} (19#2) #3}
\def\IJMP#1#2#3{{\it Int.\ J.\ Mod.\ Phys.}\/ {\bf A#1} (19#2) #3}
\def\nuvc#1#2#3{{\it Nuovo Cimento}\/ {\bf #1A} (#2) #3}
\def\RPP#1#2#3{{\it Rept.\ Prog.\ Phys.}\/ {\bf #1} (19#2) #3}
\def\etal{{\it et al\/}}

\bibliographystyle{unsrt}

\end{document}